\documentclass[a4paper,11pt]{article}
\pdfoutput=1 % only if pdf/png/jpg images are used

\usepackage{jinstpub}
\title{Construction and Assembly of the Wire Planes for the MicroBooNE Time Projection Chamber}

\author[b]{R. Acciarri}
\author[j]{C. Adams}
\author[i]{J. Asaadi}
\author[b]{J. Danaher}
\author[j]{B.T. Fleming}
\author[b]{R. Gardner}
\author[d]{S. Gollapinni}
\author[f]{R. Grosso}
\author[h,1]{R. Guenette\note{Corresponding author.}}
\author[c]{B.R. Littlejohn}
\author[b]{S. Lockwitz}
\author[b]{J.L. Raaf}
\author[e]{M. Soderberg}
\author[f]{J. St. John}
\author[b]{T. Strauss}
\author[g]{A.M. Szelc}
\author[a]{B. Yu}

\affiliation[a]{Brookhaven National Accelerator Laboratory, Upton, NY 11973, USA}
\affiliation[b]{Fermi National Accelerator Laboratory, Batavia, IL 60510, USA}
\affiliation[c]{Illinois Institute of Technology, Chicago, IL 60616, USA}
\affiliation[d]{Kansas State University, Manhattan, KS 66506-2601, USA}
\affiliation[e]{Syracuse University, Syracuse, NY 13244, USA}
\affiliation[f]{University of Cincinnati, Cincinnati, OH 45220, USA}
\affiliation[g]{University of Manchester, Manchester, M13 9PL, UK}
\affiliation[h]{University of Oxford, Oxford, OX1 3RH, UK}
\affiliation[i]{University of Texas Arlington, Arlington, TX 76019}
\affiliation[j]{Yale University, New Haven, CT 06511, USA}

\emailAdd{roxanne.guenette@physics.ox.ac.uk}

\abstract{In this paper we describe how the readout planes for the MicroBooNE Time
Projection Chamber were constructed, assembled and installed. We present the
individual wire preparation using semi-automatic winding machines and the
assembly of wire carrier boards. The details of the wire installation on the
detector frame and the tensioning of the wires are given. A strict quality
assurance plan ensured the integrity of the readout planes. The
different tests performed at all stages of construction and installation
provided crucial information to achieve the successful realization of the
MicroBooNE wire planes.}


\begin{document}
\maketitle
\flushbottom

\section{Introduction}

MicroBooNE is a 170t Liquid Argon Time Projection Chamber (LArTPC) that is deployed in the Booster Neutrino Beam (BNB) at Fermilab~\cite{ubooneTDR}. High purity liquid argon serves as the neutrino target and tracking medium for the particles produced in neutrino interactions \cite{rubbia}. The TPC active volume is 87 tons of LAr and the readout plane is constructed out of wires, reading out ionization electron signals.

\begin{figure}[h!t]
\begin{center}
\begin{tabular}{c}
\includegraphics[angle =0,width=13cm]{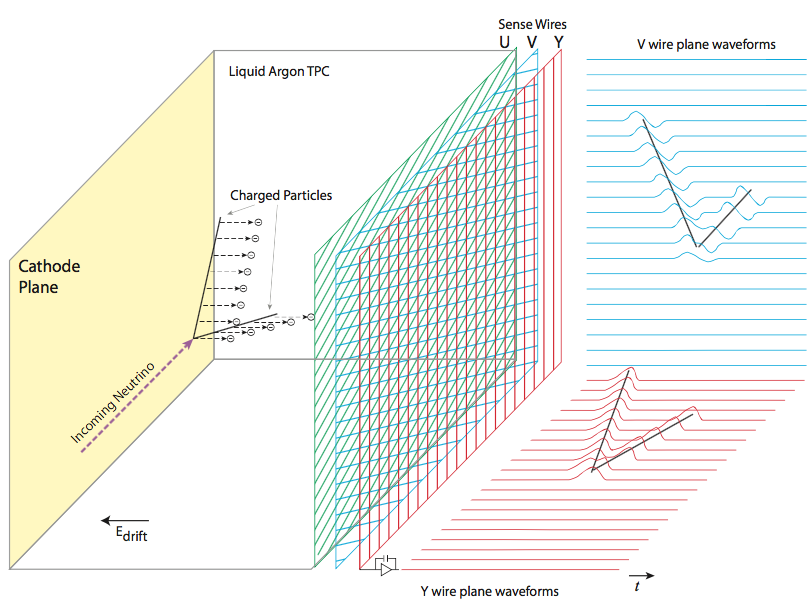}
\end{tabular}
\caption{Schematic of the operation principle of the MicroBooNE LArTPC. The charged particles leave ionization electron tracks that are drifted under an electric field towards the three wire planes. The two first planes are ``induction'' planes, where the electrons leave an induced bipolar signal as shown for the V plane. The last plane is the ``collection'' plane where the electrons are collected leaving a unipolar signal. Figure taken from \cite{uboone_detpaper}.}
\label{lartpc}
\end{center}
\end{figure}

MicroBooNE detects neutrino interactions via the outgoing charged particles (and neutral particles that have converted to charged) from the neutrino-nucleon interaction which ionize the liquid argon. Ionization electrons produced by the passage of these charged particles are made to drift at constant velocity by application of a uniform electric field. This ionization charge drifts to three wire planes situated on the beam-right side of the detector. Electrostatic potentials of the three wire planes are set up to allow the ionization electrons to pass by the first two wire planes, leaving only induced charge signals, to reach the third wire plane, where they are collected on the wires. Figure \ref{lartpc} shows the operation principle of the MicroBooNE TPC.

The successful operation of the LArTPC, and subsequent analysis of the collected data, relies on a precise knowledge of the location and orientation of all wires in the detector. Reconstruction of the charged particle trajectory is derived from the known wire positions and the arrival times of ionization electron signals on the wires, combined with the time that the interaction took place in the detector within the 1.6~$\mu$s Booster Neutrino Beam (BNB) spill. The 3~mm wire pitch, combined to the sub-millimeter resolution in the drift direction due to the MicroBooNE nominal maximum drift velocity of 1600 m$/$s and the 2 MHz sampling rate, enables 3d-position resolutions of the order of a cubic-millimeter. The full description of the MicroBooNE detector and subsystems can be found here \cite{uboone_detpaper}.

%The MicroBooNE experiment had several research and development goals to inform the design of the next generation of LAr detectors and 
\section{Description of the MicroBooNE wire planes}\label{planes}

\begin{figure}[h!t]
\begin{center}
\begin{tabular}{c c}
\includegraphics[angle =0,width=7cm]{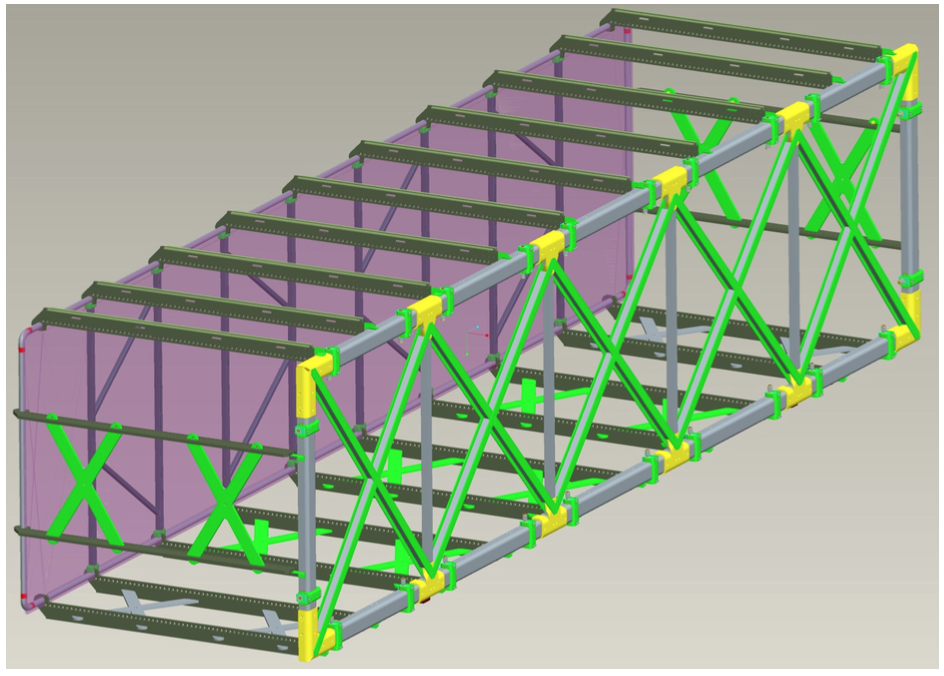}&
\includegraphics[angle =0,width=7cm]{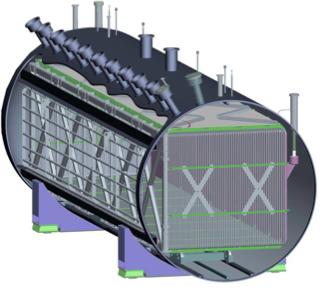}
\end{tabular}
\caption{Left: Drawing of the MicroBooNE TPC frame. The solid cathode is located on the left side and the anode frame is visible on the right side. The wire planes attach to the anode frame. Right: Drawing of the TPC frame inserted in the cryostat. In this drawing, the wire planes are locate on the left side. Figures taken from \cite{uboone_detpaper}.}
\label{tpc_frame}
\end{center}
\end{figure}

The MicroBooNE TPC design is shown in figure \ref{tpc_frame}. The TPC, with dimensions of 2.56~m~(drift length from cathode to anode) $\times$~2.33~m~(height of the readout plane) $\times$~10.36 m (length of the readout plane), is inserted in a cylindrical cryostat. The active volume is defined as the volume that is readout by the wire planes, hence inside the TPC frame.

\begin{figure}[h!t]
\begin{center}
\begin{tabular}{c}
\includegraphics[angle =0,width=12cm]{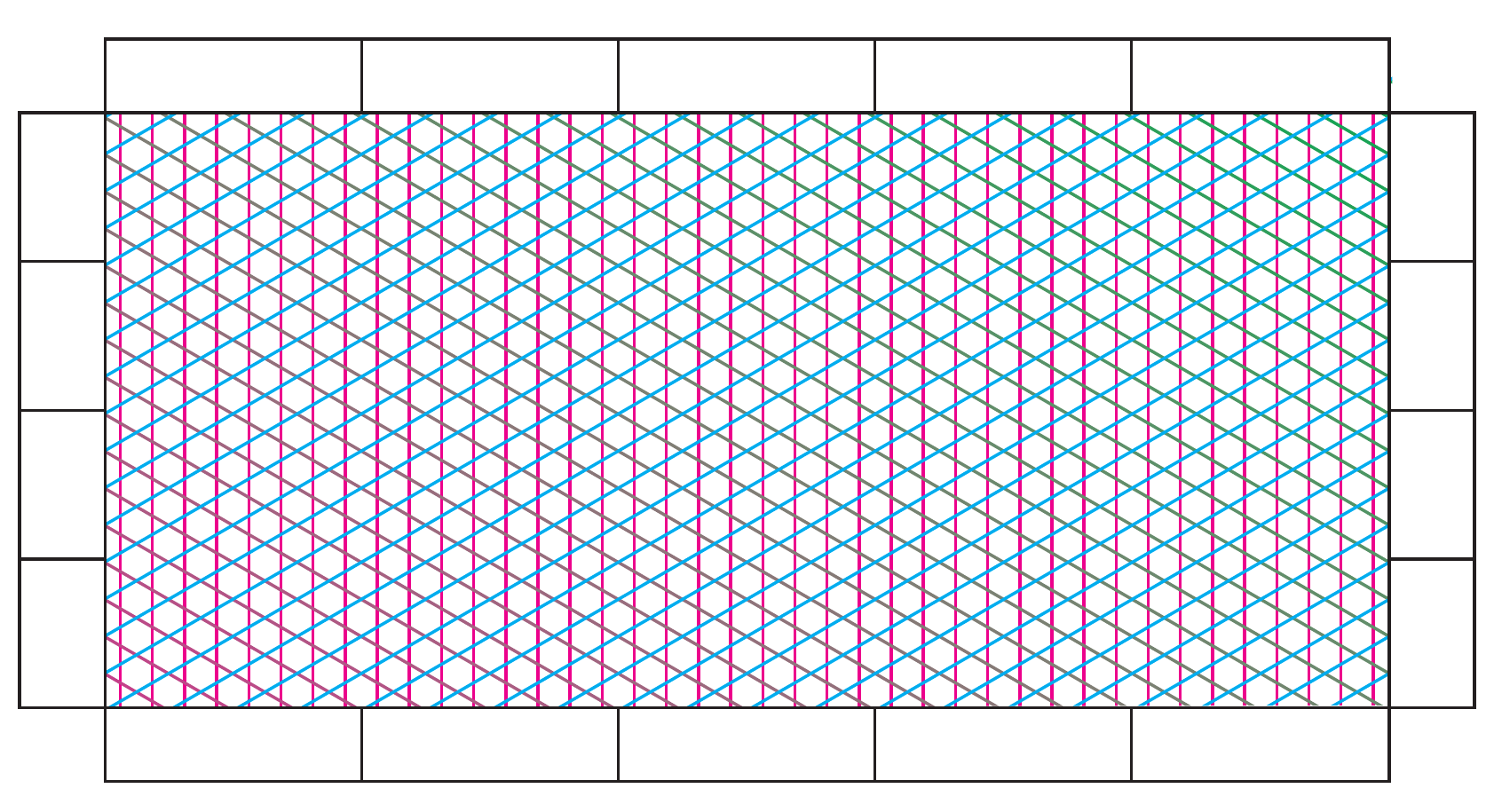}
\end{tabular}
\caption{Simplified schematic view of the wire planes. The vertical Y wires are 
shown in pink, the U wires, angled at +60$^{\circ}$ are shown in blue and the V 
wires, angled at -60$^{\circ}$, are shown in green. The illustration is not to scale, nor are the boards included in the drawing.  It is for illustrative purposes only}
\label{schema_planes}
\end{center}
\end{figure}

The design of the MicroBooNE wire planes is based partly on the ICARUS
detector \cite{icarus}. The TPC is composed of three wire planes Y, U and V,
oriented vertically and at $\pm$ 60$^{\circ}$ from the vertical,
respectively, as illustrated in figure \ref{schema_planes}. The planes are 2.33 m $\times$~10.36 m, the plane-to-plane
separation is 3~mm, and the wire-to-wire separation is also 3~mm. The
vertical Y-plane, which serves as the collection plane, is comprised of 3456
individual wires mounted in groups of 32 on wire carrier boards, for a total
of 108 pairs of wire carrier boards installed for the Y plane. The wire carrier boards
are described in greater detail in section~\ref{sec_boards}. The two angled
planes, U and V, serve as the two induction planes and are each comprised of
2400 wires, mounted in groups of 16 on 150 pairs of carrier boards. 

The wires are made of $150\pm 5$ $\mu$m diameter Type-304V stainless steel, the same material as the TPC frame to ensure identical thermal contraction. The wires are
plated with 2~$\mu$m of copper (to decrease the resistivity of the wire from 40 $\Omega /$m to 3 $\Omega /$m in
order to reduce series noise in the readout electronics \cite{copper}) and covered with a
thin flash (0.1~$\mu$m) of gold to prevent the copper oxidation. %During
%detector operation, appropriate electric fields are maintained to prevent
%collection of electrons on the U and V wires. At the nominal electric field of 500
%V/cm, the bias voltages of -204~V for the U plane, 0~V for the V plane
%and +440V for the Y plane will achieve maximal transmission of electrons
%through the first two induction planes.

%\begin{figure}[h!t]
%\begin{center}
%\begin{tabular}{c}
%\includegraphics[angle =0,width=12cm]{Field_Map.pdf}
%\end{tabular}
%\caption{}
%\label{Emap}
%\end{center}
%\end{figure}

% This view is taken from a 3D simulation of the MicroBooNE wire planes, with the nominal bias voltages on each plane.  This view is a cut plane perpendicular to the Y plane, cutting the U & V wires at 60degree angles and resulted in the doubling of the wire pitch.  The color contours is the electrical potential, and the electric field lines are in black.

The original technical requirements for the wire plane design are fully detailed here \cite{ubooneTDR}. Three wire planes were required to ensure redundancy to allow reconstruction of tracks parallel to one wire plane. The wire pitch and the wire plane separation were chosen to be 3~mm, the minimum spacing possible given the electronics signal-to-noise ratio and LAr purity requirements. The wire tension was originally designed to be 9.8~N to limit the sag due to gravity to be less than 0.5~mm on the longest 4.66~m wires. The tolerance on the length was $\pm$ 0.02\%. However, it is important to mention that due to the R\&D goals of MicroBooNE, some original requirements were modified during the final design and construction processes due to physical constraints or because it was demonstrated by simulations that the physics goals could still be achieved. For example, the final requirement on the tension in the TPC is 6.86~N. However, the tensioning device design did not allow to tension all the individual wire tensions within a factor two of that desired value. More discussion on these requirement changes can be found in section \ref{tension}.

\subsection{The wire carrier boards}\label{sec_boards}
\begin{figure}[htpb!]{}
\begin{center}{}
\begin{tabular}{ccc}{}
\includegraphics[angle =0,width=13.5cm]{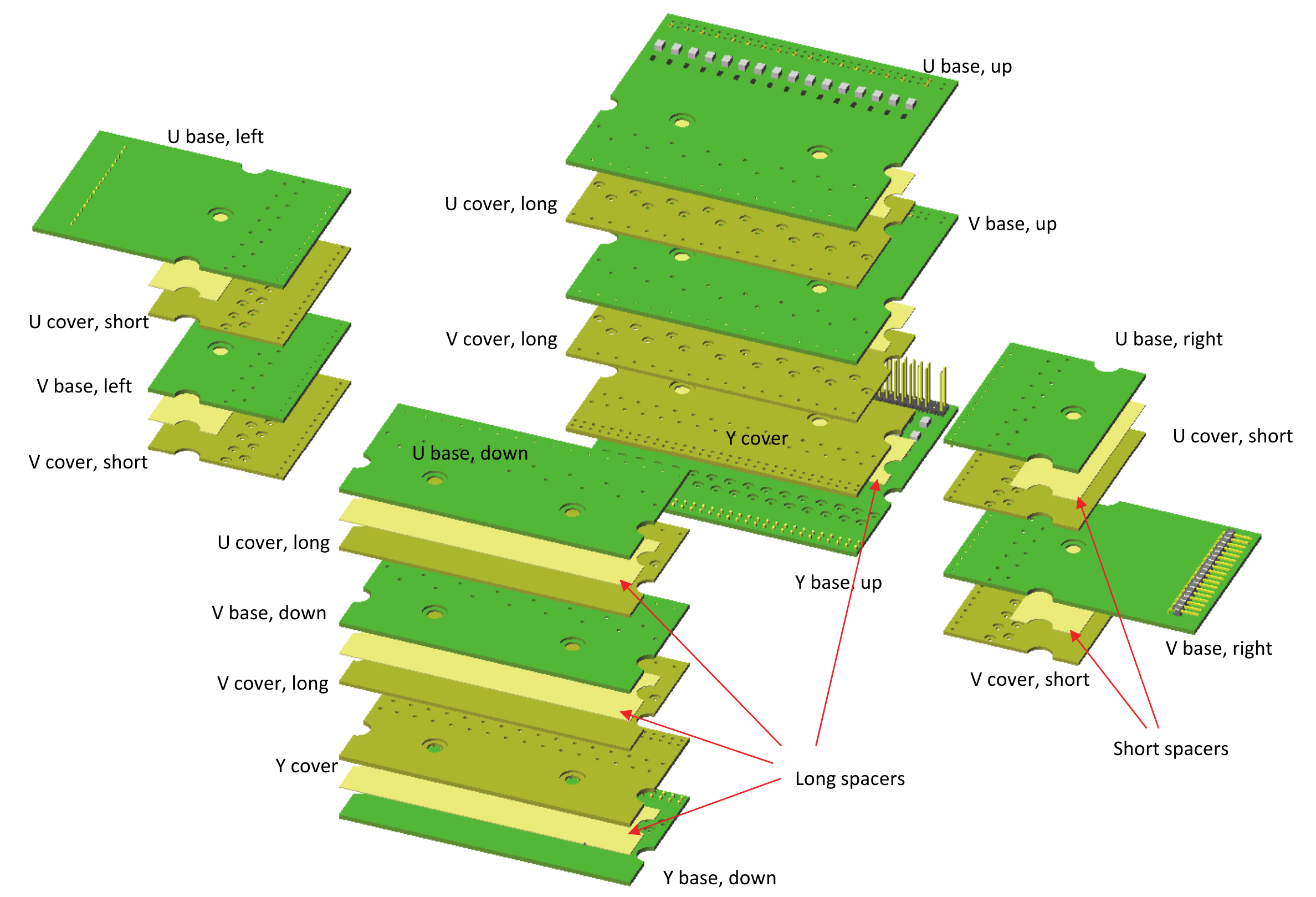}&
\end{tabular}
\caption{Exploded-view schematic of the wire carrier board assemblies for the three
planes. This drawing shows how the boards stack once installed. Each board
assembly consists of a base (where the wires are individually inserted), a
spacer and a cover (that secures the wires in place and is closed using
rivets in the central holes).}
\label{board_assembly}
\end{center}
\end{figure}

The wire carrier boards, shown schematically in figure~\ref{board_assembly},
are made of FR4, comprised of three parts: a base, a cover, and a spacer,
which are made using standard printed circuit board fabrication techniques.
Rows of counter-bored pockets are machined into the base and cover.  Once
assembled, they form pockets that securely capture the brass wire-termination
rings, described in section~\ref{winding}. The spacer maintains a small gap
between the leading edge of the base and cover so the wires can move slightly
around the terminating brass rings.  Electrical contact to each wire is made
by gold plated brass pins pressed in at the leading edge of the base board.
 These pins also define the location of the wires as
they exit the wire carrier modules. The wires are hence attached to two boards: one to connect the wires to the detector readout system (top boards for the Y plane, left and top boards for the U plane and right and top boards for the V plane) and one, non-instrumented, to terminate and secure the wires on the frame (bottom boards for the Y plane, right and bottom boards for the U plane and left and bottom boards for the V plane) as shown in figure \ref{board_assembly}.

\section{Wire winding and carrier board assembly}\label{winding}

\begin{figure}[htbp!]
%\begin{center}
\begin{tabular}{cc}{}
\includegraphics[angle =0,width=8.5cm]{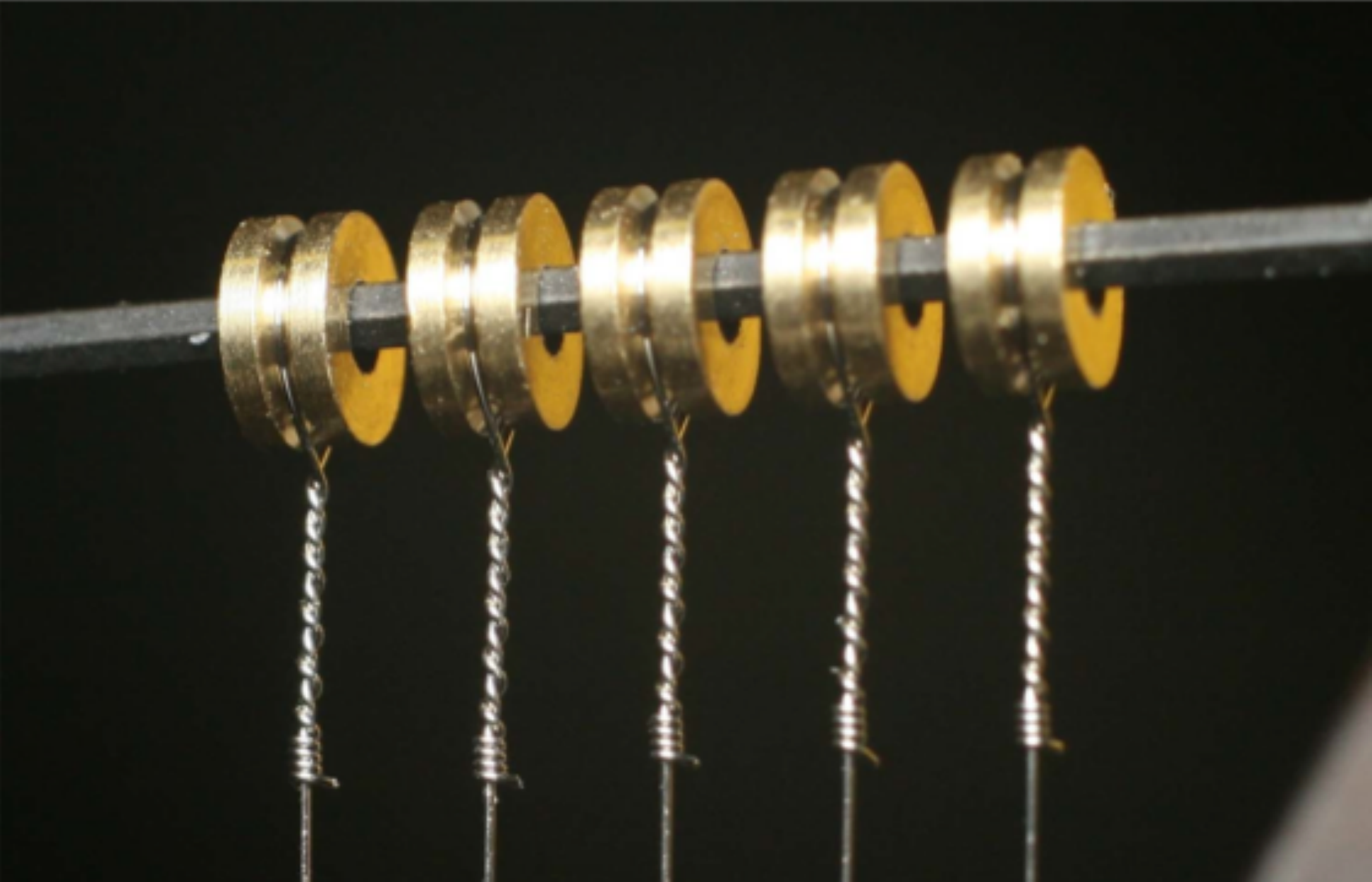}&
\includegraphics[angle =0,width=7.4cm]{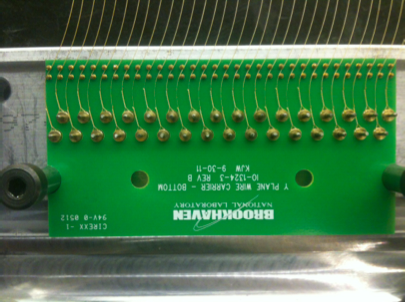}\\
 \hspace*{0.2cm}
\includegraphics[angle =0,width=8.4cm]{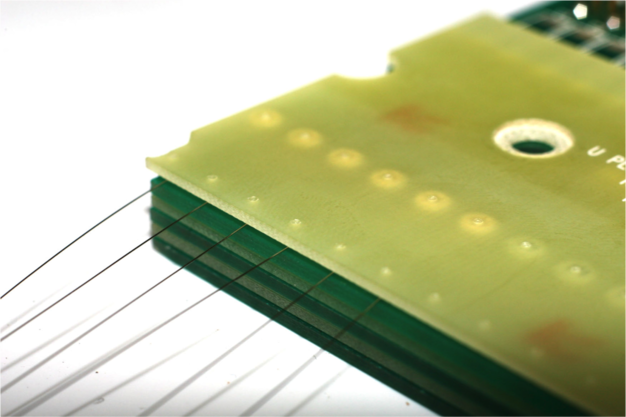}
\end{tabular}
\caption{Top left: Picture of the wire termination on the brass rings. Top right: Example
carrier board. The two rows of pressed-in pins that guide the wires and
maintain the 3~mm spacing are visible at the top of the board. These pins
also ensure the electrical connections between the wires and the board
electronics. Bottom: Example of a completed wire carrier module before rivets
have been placed in the large central holes (only one shown here) to hold the
module securely closed.}
\label{board}
%\end{center}
\end{figure}

In order to construct the MicroBooNE wire planes, each TPC wire is prepared
(cut to length and twisted around a brass ring as shown on the left in
figure~\ref{board} and attached individually on a wire carrier board (center
of figure~\ref{board}). Once the wire carrier board is fully populated, a
board spacer is inserted and the board is closed by a cover, and secured with
two rivets (partial view shown in right panel of figure~\ref{board}, shown
without rivets). The preparation of each individual wire is facilitated by the
use of automated wire winding machines designed for this purpose. In this
section, we describe the wire winding machines and the winding procedures.

\subsection{Wire winding machines}\label{sec:y3}

The winding of MicroBooNE's wires is performed via a semi-automated procedure
using wire winding machines designed to measure and cut the desired wire
lengths and to consistently wind the wires on the wire termination brass
ring. Two machines were designed and built. One is used for fixed-length
wires (for the vertical Y plane) and the other for variable-length wires (for
the angled U and V planes).

\begin{figure}[!th]{}
\begin{center}{}
\begin{tabular}{c}{}
\includegraphics[angle =0,width=15.5cm]{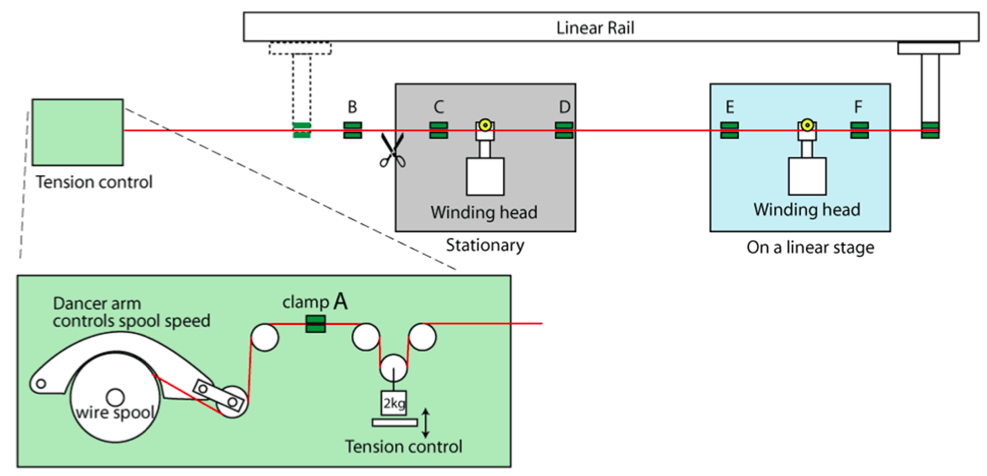}
\end{tabular}
\caption{Simplified schematic of the wire winding machine. On the left, the
distribution station where the wire is unrolled and adjusted to the desired
tension. The linear translation stage shown at the top is used to measure and
cut the wires to precise lengths. Two winding heads, described in more detail below, are also shown that wrap
the wire ends around a brass ring. The letters A to F represent clamps that hold the wire in place during operation.}
\label{machine}
\end{center}
\end{figure}

Figure \ref{machine} shows a schematic view of a wire winding machine. The
first component of the machine is the distribution and tension station.  A
spool of wire is attached to the machine frame and the wire is unrolled with
a robotic arm and pulley system. A weight of 1.4~kg is suspended from a
mobile pulley in order to apply the nominal 0.7~kg or 6.86~N tension on the wire. This
procedure ensures that the wires are cut to the correct length under tension
at room temperature. The nominal tension of 6.86~N was chosen to be small enough to
prevent wire breakage during cooldown, and large enough to limit the maximum
wire sag due to gravity to be less than 0.5~mm.

%Figure \ref{tensionstation} shows a close-up picture of these stations.
%\begin{figure}[!ht]{}
%\begin{center}{}
%\begin{tabular}{c}{}
%\includegraphics[angle =0,width=8.5cm]{tensionstation.png}
%\end{tabular}
%\caption{Picture of the distribution and tension stations of the MicroBooNE wire winding machine. The wire is unrolled from the spool with the rotating arm in top of the spool and passed through the different pulleys. The clamped at the top is used to maintained the wire in place and the weight is released when the wire is cut to the desired length to apply to appropriate tension on the wire.}
%\label{tensionstation}
%\end{center}
%\end{figure}

\begin{figure}[ht!]{}
\begin{center}{}
\begin{tabular}{c}{}
\includegraphics[angle =0,width=10.5cm]{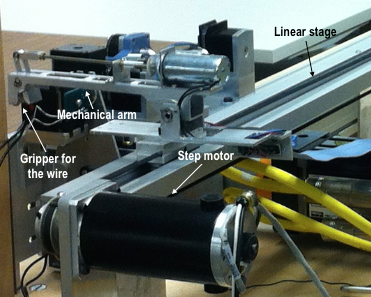}
\end{tabular}
\caption{Picture of the mechanical arm with gripper (left side of image, near top)
mounted on the linear stage. The arm grabs the wire and translates along the
linear stage to pull it to the desired length, controlled by the step motor
(black, at bottom of image).}
\label{arm}
\end{center}
\end{figure}

The second part of the machine is composed of a robotic arm mounted on a
linear translation stage, as shown in figure~\ref{arm}. The arm is used to
pick up the wire from the distribution station and pull it to the desired
length. The linear stage provides excellent control on the length (better than 50 $\mu$m precision) and is operated by computer controlled step motor. The Y plane has a fixed length of 2.5~m for all wires, but the U and V planes have
variable wire lengths from 3.7~cm (in the corners) up to 4.66~m due to their 60$^{\circ}$ orientation and position along the frame.

\begin{figure}[!ht]
\begin{center}
\begin{tabular}{c}
\includegraphics[angle =0,width=14.5cm]{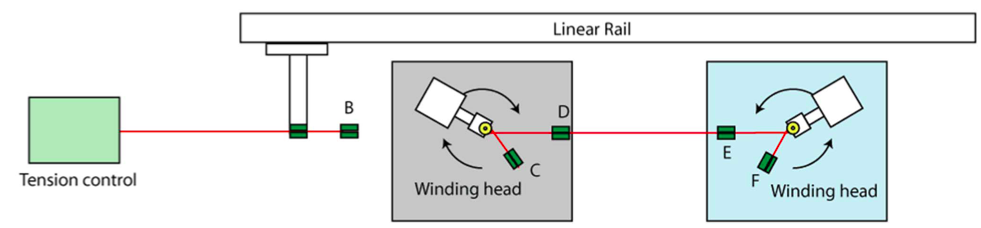}\\
\includegraphics[angle =0,width=5.5cm]{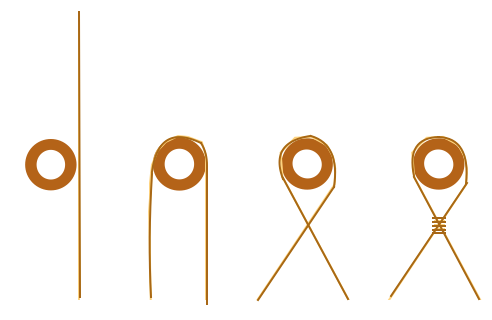}\\
\includegraphics[angle =0,width=8.0cm]{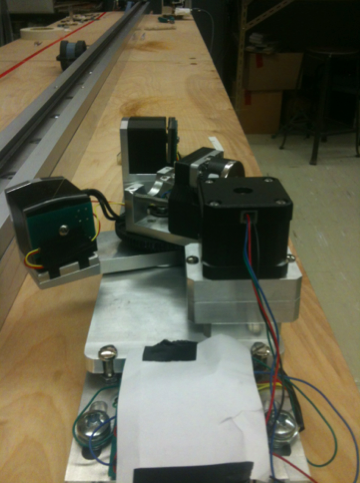}
\end{tabular}
\caption{Top: Schematic of the winding heads. Middle: Schematic of the winding steps
around the brass rings (1. The wire is deposited behind the ring. 2. The
winding head wraps the wire around the ring. 3. The head completes the
rotation, bringing the wire to the desired angle for the twist. 4. The winding
head rotates in the plane perpendicular to the previous rotation, twisting the
wire to secure it on the ring). Bottom: Photograph of one of the winding
stations.}
\label{heads}
\end{center}
\end{figure}

Finally, the third part of the machine consists of two winding stations (see
figure~\ref{heads}) designed to attach each end of the wires to a small brass
ring. The brass rings are 3~mm in diameter and 1.5~mm thick. The robotic arm
deposits the wire just behind the brass rings (pre-installed manually on the
winding heads) where it is clamped in place by electrically actuated
grippers, and cutters automatically cut the wire. Once the wire is released
by the arm, the winding heads rotate by 120$^{\circ}$ and start spinning in
the plane perpendicular to the initial rotation, winding the wire around the
ring six times. A twisted wire on the ring is shown on the top left in
figure~\ref{board}. Once the winding is performed, the machine's
work, the automated part of the procedure, is complete. It goes to rest and
operators take over to complete the procedures described in the next
section.

\subsection{Procedures for the assembly of the wire carrier boards}\label{sec:y4}

After the automated stage, the wire (of the correct length) is left twisted
around the brass rings. The final step of the wire preparation requires an
operator to manually cut the the excess wire remaining at the end of the
twisted wire (cutting away the wire section from the brass ring to block C,
as well as the section from the other brass ring to block F, both shown in
figure~\ref{heads}), ensuring that the cut is very close to the twist to
avoid sharp edges and short-circuits in the boards.

\begin{figure}[!ht]{}
\begin{center}{}
\begin{tabular}{c}{}
\includegraphics[angle =0,width=13.5cm]{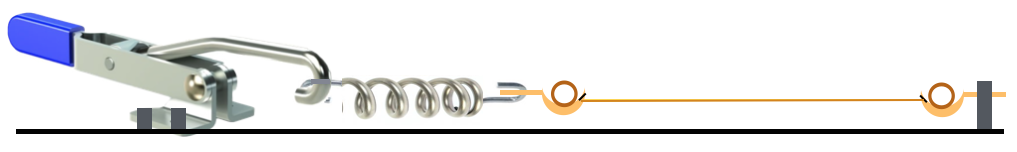}\\
\includegraphics[angle =0,width=14.0cm]{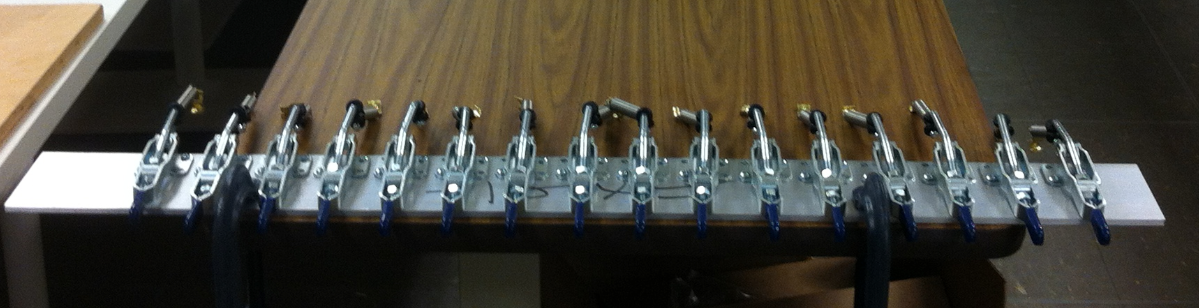}
\end{tabular}
\caption{
Top: Schematic of the wire tension device used to test the wires. Bottom:
Photograph of the actual individual wire tension station where a tension of
24.5~N is applied to every wire for 10 minutes. The wire is secured to a
spring mounted on a lever. Once the lever is lowered, the spring elongation
creates a force equivalent to 24.5~N.}
\label{ind_tension_station}
\end{center}
\end{figure}

Once a wire has been prepared, the operators manually remove it from the
machine by lifting the rings from the winding machine's pins. The wire is
then strength-tested. Each individual wire is placed on a tensioning station,
where a tension of 24.5~N (2.5~kg) (more than three times the nominal TPC tension) is
applied for ten minutes. This ensures that the wire preparation has not
caused any weaknesses that could result in a breakage later
on. Figure~\ref{ind_tension_station} shows the wire tension testing station,
where springs mounted on levers create the force equivalent to 24.5~N on
the wires when the lever is completely pulled down. One extremity of the wire
is attached to the spring and the other one is fixed to the station.

\begin{figure}[!ht]{}
\begin{center}{}
\begin{tabular}{c}{}
\includegraphics[angle =0,width=9.5cm]{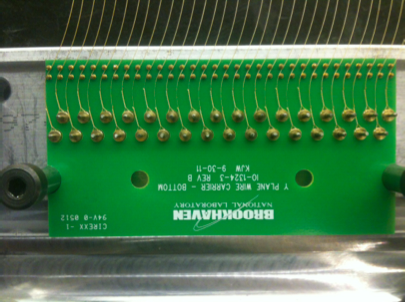}
\end{tabular}
\caption{Photograph of a Y plane wire board manually filled with the 32 prepared wires.}
\label{carrier_boards}
\end{center}
\end{figure}

Having controlled the quality (strength) of each wire, the wires are
placed manually in wire carrier modules shown in
figure \ref{carrier_boards}. To ensure good electrical contact to the wires, two rows of pins are used for the Y board, where each wire makes an $S$-shaped bend between two pins (shown
in figure~\ref{board}). A single row of pins are used on the U $\&$ V boards,
since the wires are pulled against the pins at a 60 degree angle which
ensures good contact. The board spacer and
cover are installed and the board is closed. The assembled board is then placed in a stand for
stress-testing, and a tension of 24.5~N per wire is reapplied to the fully
populated board for 10 min. to ensure that the wires have not been
weakened during the board assembly process. Finally, the assembled wire boards are placed into custom-made storage boxes where they remain until their installation on the TPC frame.

%The holes (pockets) in the carrier boards, which
%are the same size as the brass rings, ensure that the wires are secured
%inside the boards. Once the board is fully populated, the board spacer and
%cover are installed and the module is sealed by holding rings (rivets)
%pressed on the board. The assembled board is then placed in a stand for
%stress-testing, and a tension of 2.5~kg per wire is reapplied to the fully
%populated board for 10 min. This is to ensure that the wires have not been
%weakened during the board assembly process. The stress stand is shown in
%figure~\ref{stress_stand}, where a design similar to that of the individual
%wire tension station is shown. Strong springs are attached to a bigger lever
%that is pulled down to apply a force of 80~kg for the Y plane boards and 40~kg
%for the U and V plane boards.

%\begin{figure}[tbpht!]
%\begin{center}
%\begin{tabular}{c}
%\includegraphics[angle =0,width=7.0cm]{tension.png}
%\end{tabular}
%\caption{Photograph of the board tensioning station where springs are mounted to a lever applying the tension to the whole board.}
%\label{stress_stand}
%\end{center}
%\end{figure}

%\begin{figure}[tbpht!]
%\begin{center}
%\begin{tabular}{c}
%\includegraphics[angle =0,width=7.0cm]{box.jpg}
%\end{tabular}
%\caption{Photograph of the custom-made box for U and V plane wire storage and
%transport where the boards are stacked on top of each other at a 60$^{\circ}$
%angle.}
%\label{boxes}
%\end{center}
%\end{figure}

%\begin{figure}[tbp]{}
%\begin{center}{}
%\begin{tabular}{c}{}
%\includegraphics[angle =0,width=15.0cm]{layout.png}
%\end{tabular}
%\caption{
%Schematic of the layout for the full wire preparation operation. From left to
%right, the winding machines, followed by the tension stations and the storage
%box are illustrated.}
%\label{lab_setup}
%\end{center}
%\end{figure}

\section{Tests and quality controls of the wires and the boards}\label{tests}

One of the biggest risks to a TPC is that a wire might break after
installation in a sealed cryostat filled with liquid argon. If even one of
the 8256 wires happens to break once installed, it could short-circuit a
significant portion of the signal wires if the broken wire makes contact
with neighboring wires. In the worst case scenario, a broken wire could
short-circuit the entire TPC if it were to make contact with the cathode plane
where the high voltage is applied. Re-opening the cryostat to repair a broken
wire would be both time-consuming and expensive, causing serious problems to
the experiment. In order to ensure that no wires break after their
installation, a series of tests are performed to understand the wire
limitations. 

Every aspect of the wire preparation was previously studied carefully to
optimize the quality of the wires. The wire diameter, number of twists on the
brass rings, the twisting angle and the post-preparation tests were all
studied in great detail. The different tests and studies are presented in
this section.

\subsection{Wire strength and breakloads}
It was found that 150~$\mu$m wire diameter was sufficient to safely withstand the
TPC nominal tension of 6.86~N. The wire manufacturer reports a
breakload of 42.83~N, which is over six times the TPC nominal tension. Tests to verify the wire breakload are performed and the results for 20 wires are shown in figure \ref{summary_breaks}. At room temperature, the wire breakload measured of 42.9 $\pm$ 1.5 N is in agreement with the vendor specifications. At liquid nitrogen (LN$_{2}$) temperature, four wires were tested and the measured breakload of 50.4 $\pm$ 0.9 N is over seven times the nominal tension.

%\begin{figure}[ht!]{}
%\begin{center}{}
%\begin{tabular}{c}{}
%\includegraphics[angle =0,width=12cm]{break_rt.pdf}
%\end{tabular}
%\caption{Breakload for 20 wires at room temperature.}
%\label{break_rt}
%\end{center}
%\end{figure}

\subsection{Wire strength and breakloads of damaged wires}

Defects or damages on the wires may reduce the breakload considerably. One
type of damage that can occur is if the wire gets kinked. To test the effect of kinks on the breakloads, single wires are kinked using pins of three different diameters (0.15~mm, 0.53~mm and 1~mm). The
range of diameters tests the effect of sharper kinks. The results of these
tests on three wires at room temperature and one wire at LN$_{2}$ are summarized in figure~\ref{summary_breaks}. The breakload decreases by as much as 20\% for small-diameter kinks. As expected, the sharper the kink, the greater the reduction in breakload. The breakload of damaged wires is also higher
in LN$_{2}$ than at room temperature. In MicroBooNE, visual inspection was in place at several stages of the procedures to remove any kinked wires, therefore these extreme cases should not be present in the TPC. However, if several visual inspections missed a kinked wire, these results show that the safety margin was above five times the nominal tension at room temperature and above six times the nominal tension at LN$_{2}$. \\

%\begin{figure}[!ht]{}
%\begin{center}{}
%\begin{tabular}{c}{}
%\includegraphics[angle =0,width=3.7cm]{kink1.pdf}
%\includegraphics[angle =0,width=5.2cm]{kink2.pdf}
%\end{tabular}
%\caption{Examples of artificial kinks introduced on wire using pins of differing diameters.}
%\label{kink}
%\end{center}
%\end{figure} 

\begin{figure}[ht!]{}
\begin{center}{}
\begin{tabular}{c}{}
\includegraphics[angle =0,width=9.5cm]{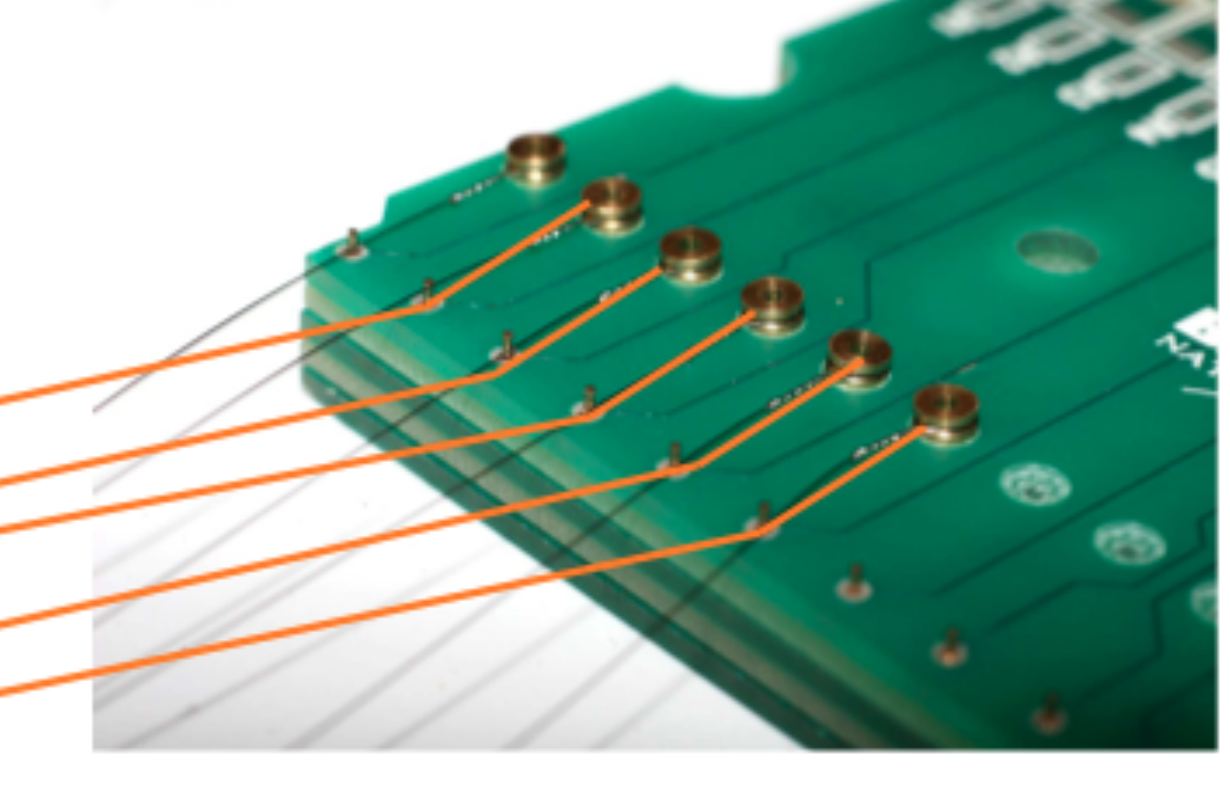}
\end{tabular}
\caption{
Example of how the wires (in orange) of the U and V planes are bent around
the pins in the wire carrier boards. Once installed, these wires are angled
at 60$^{\circ}$ from vertical.}
\label{bentwires}
\end{center}
\end{figure}

Another effect to consider when studying wire breakage is the effect of
bending wires. For the U and V planes of the TPC, the wires are bent around
1~mm pins on the wire carrier boards, angled at 60$^{\circ}$ once installed
in the TPC (see figure~\ref{bentwires}). Tests are performed to understand if
bending a wire around a small object may reduce its breakload. To perform
these tests, single wires are bent at 60$^{\circ}$ around pins of 1~mm and
1.2~mm in a way similar to how they are bent around the pins on the wire
carrier boards for both room and LN$_{2}$ temperatures. The results of these tests are also summarized in
figure~\ref{summary_breaks}. As expected, the breakload decreases with the pin diameter and a bent has a less dramatic effect than a kink for the wire strength. For the 1~mm pin used in MicroBooNE, the breakload measured was over five times higher than the nominal tension at room temperature and near 6.5 times the nominal tension at LN$_{2}$ temperature.

\subsection{Effects of wire termination and attachment}
\begin{figure}[ht!]{}
\begin{center}{}
\begin{tabular}{c}{}
\includegraphics[angle =0,width=6.0cm]{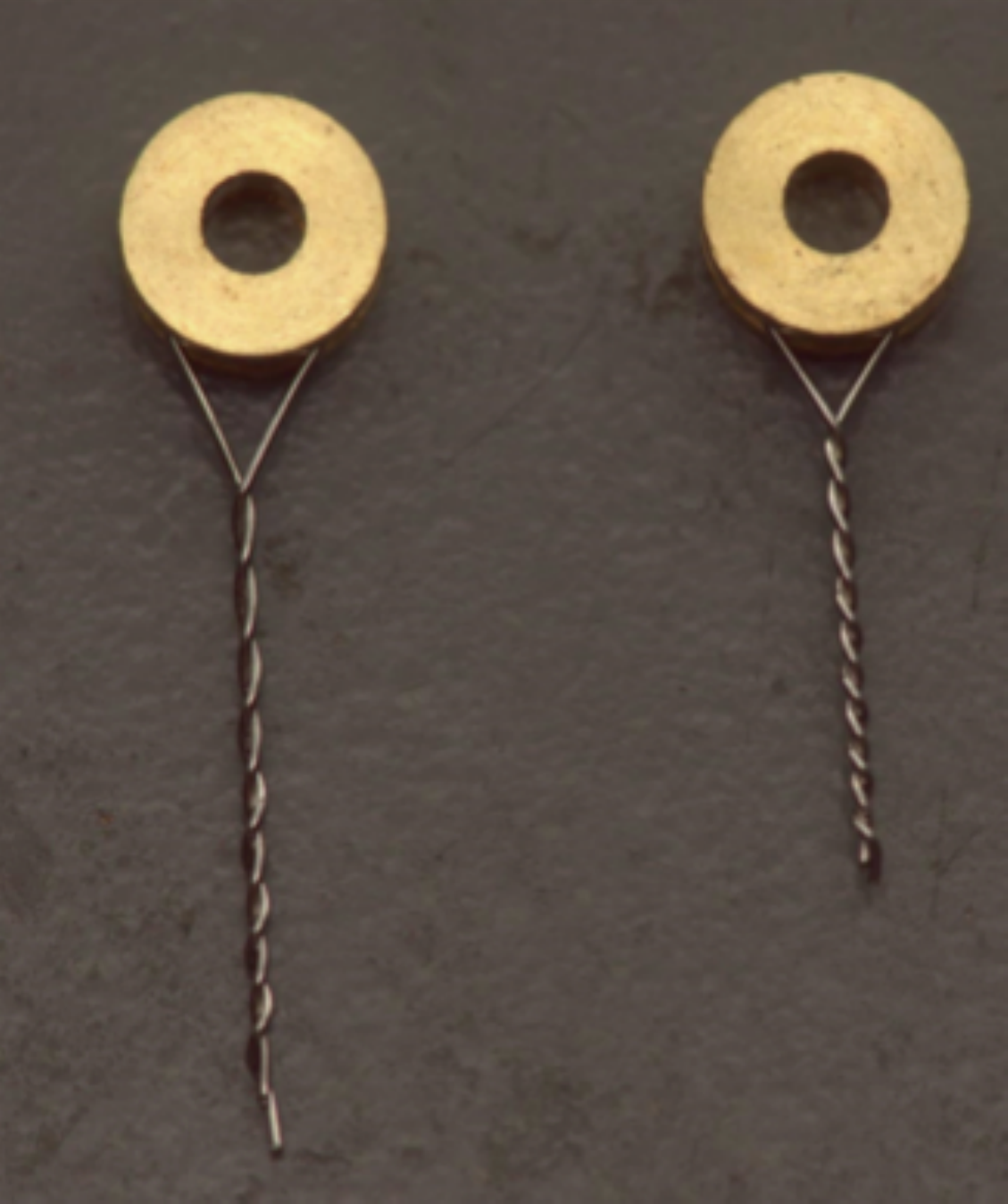}
\end{tabular}
\caption{Two twist angles around the brass rings. Left: Angle of 50$^{\circ}$. Right: Angle of 60$^{\circ}$.}
\label{twist}
\end{center}
\end{figure}

\begin{figure}[tbp]{}
\begin{center}{}
\begin{tabular}{c}{}
\includegraphics[angle =0,width=15.0cm]{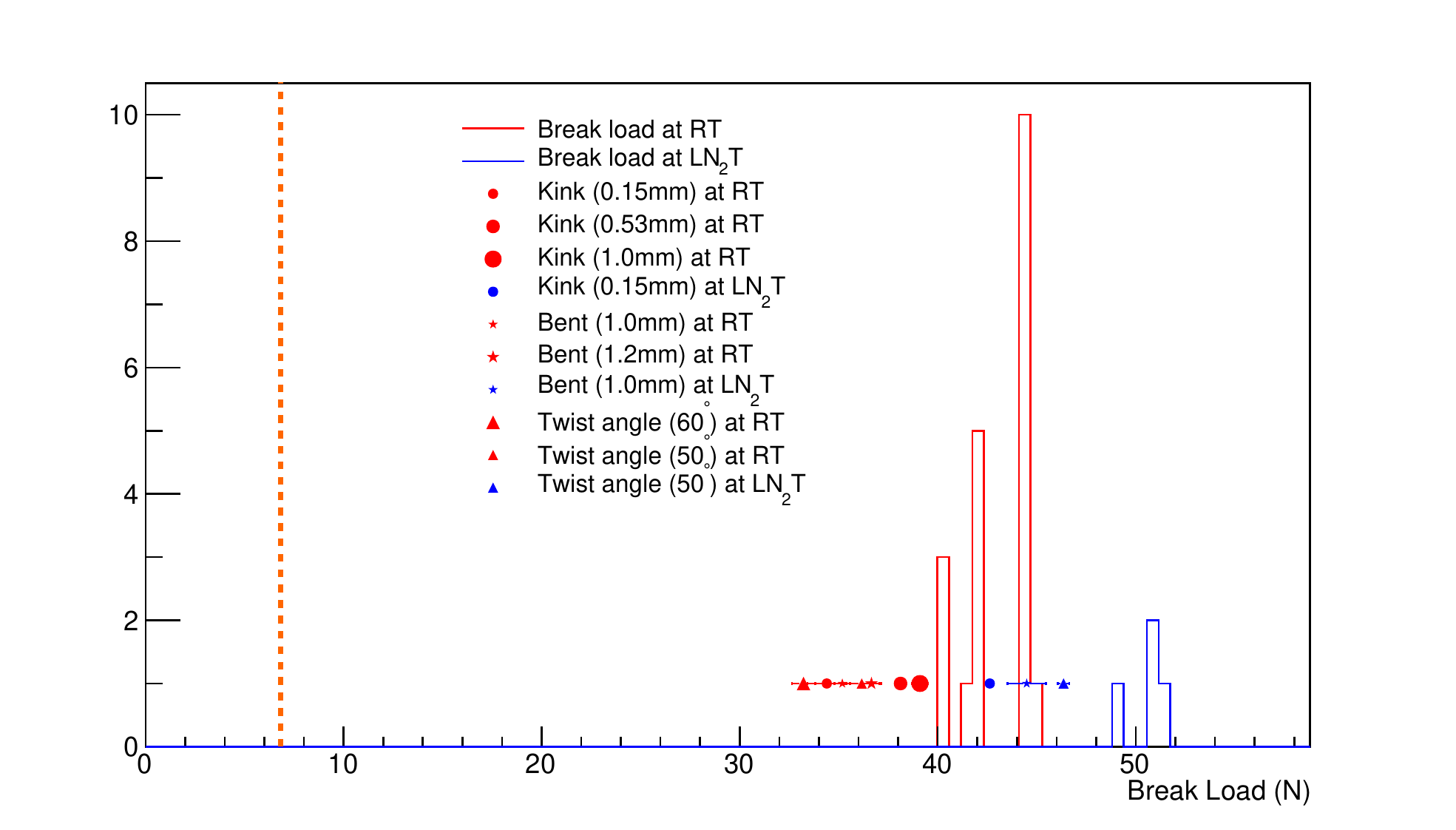}
\end{tabular}
\caption{Summary of all breakload tests performed. The data points represent individual measurements and the histograms, sets of multiple measurements. The red points and histogram represent tests done at room temperature (RT) and the blue points and histogram represent the tests done at LN$_{2}$ temperature. The orange dashed
line shows the nominal tension that is used in the TPC.}
\label{summary_breaks}
\end{center}
\end{figure}

The winding of the wires around the brass rings is also studied in order to
optimize the number of twists and the twist angle. It is found that 6 twists
are satisfactory to combine maximal strength and minimal length inside the carrier
board. The angle at which the winding machines
twist the wires is investigated; figure~\ref{twist} shows two different
twist angles of 50$^{\circ}$ and 60$^{\circ}$. The twisting angle tests results of two individual wires at room temperature and one wire at LN$_{2}$ temperature are shown in figure~\ref{summary_breaks}. The  10\% increase in breakload going from 60$^{\circ}$ and 50$^{\circ}$ at room temperature is over five times the nominal tension and over 6.5 times for LN$_{2}$ temperature. The 50$^{\circ}$ angle was therefore used for MicroBooNE. 

\subsection{Effect of cold temperature shock}
A final test is performed to see the effect of a very rapid change from room
temperature to LN$_{2}$ temperature on a fully-populated wire carrier
board. This test uses pre-assembled Y plane wire boards placed under
different tensions equal to or above the nominal detector tension. The tensioned
(or over-tensioned) boards are then plunged into a LN$_{2}$ bath to test if any
wires will break. The different tensions  applied to the wire assemblies during the test are 6.86N/wire (nominal TPC tension), 10.4N/wire, 12.4N/wire, 10N/wire and 18 N/wire (over 2.5 times the nominal tension). No wire breakage is observed for any of the applied tensions.

%\begin{figure}[tbp]{}
%\begin{center}{}
%\begin{tabular}{c}{}
%\includegraphics[angle =0,width=13.5cm]{ColdTestResults.png}
%\end{tabular}
%\caption{Summary of all overload tests where the completed wire carrier board is
%loaded with more than the nominal tension before being plunged into a LN2
%bath. No failure of either the wires or the wire carrier boards is observed
%during testing.}
%\label{fig:coldshock}
%\end{center}
%\end{figure} 

\subsection{Discussion of the tests performed} 
The tests show that extreme conditions applied to the wires reduce the
breakload. However, the minimum breakload observed in all of these tests is
33.3~N at room temperature. This tension is still almost five times
higher than the nominal tension applied in the TPC (6.86~N). In LN$_{2}$ (77K), the lowest breakload
observed is 42.63~N, which is more than six times the nominal tension. Since the TPC is immersed in LAr (87K), it is therefore clear that the safety margin is sufficient to prevent wire breakage, even in the case of
unnoticed damage to a wire. The quality assurance procedures described earlier ensure that all wires and wire carrier boards are systematically tested and inspected before installation, greatly reducing the risk of dramatic damage to the wires. The decision to sample only a small number of wires for these tests was made partly on experience from the previous ICARUS detector \cite{icarus}, that did not observed wire breakage after the detector assembly (over ten years) and from MicroBooNE collaborators handling wires in many different conditions without observing breakage.

\section{Wire installation}\label{sec:x3}

\begin{figure}[h!t]{}
   \begin{center}{}
   \begin{tabular}{c}{}
     \includegraphics[width=9.5cm]{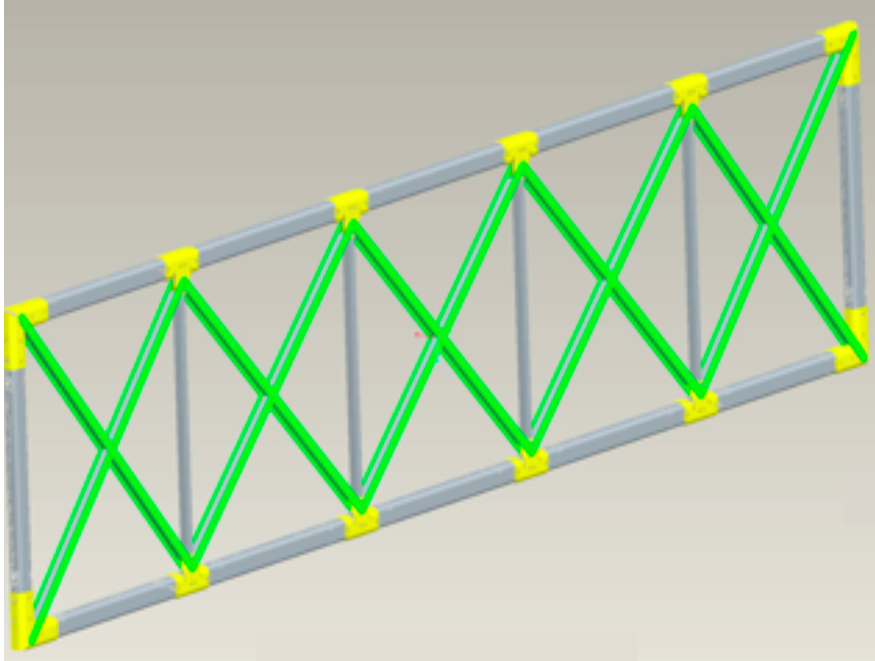}\\
     \includegraphics[width=14.5cm]{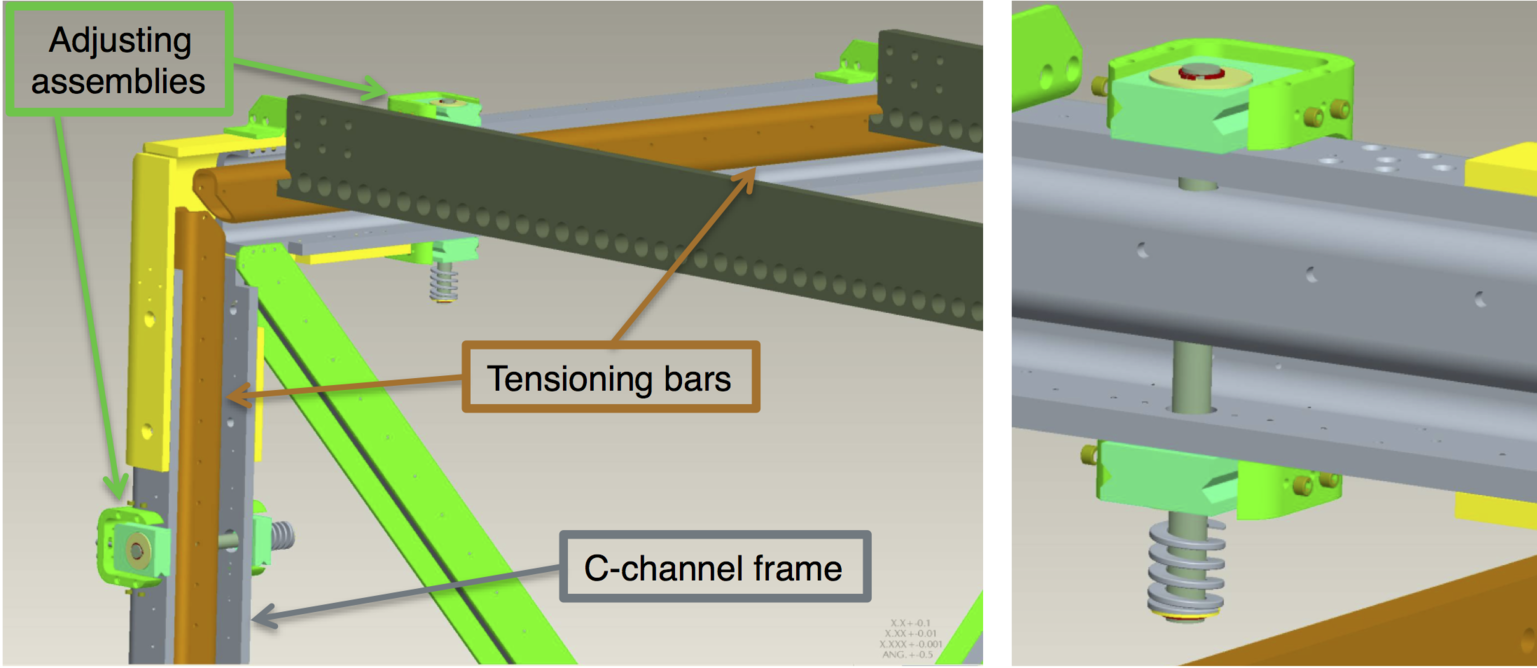}
     \end{tabular}
     \caption{Top: Drawing of the full frame for the wire plane. Bottom left: Renderings of one corner of the anode frame (top left part of the frame). Bottom right: Rendering of a close-up of adjusting assembly. The tensioning bar position (pitch, roll,
     yaw) is controlled by the shaft that extends through the adjusting
     assembly, C-channel, and tensioning bar and by tensioning screws (not shown). 
     The holes in the tensioning bar are for precision mounting pins onto which 
     the wire carrier boards are attached.}
     \label{fig:anodeframeCAD}
  \end{center}
\end{figure}

\begin{figure}[h!t]
   \begin{center}
   \begin{tabular}{c}
     \includegraphics[width=11.5cm]{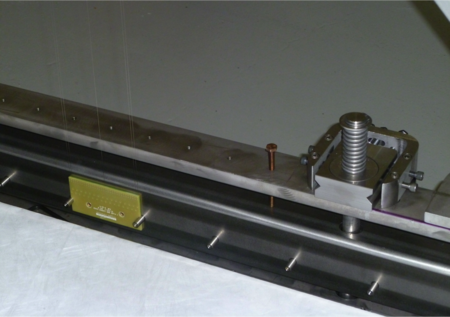}
     \end{tabular}
     \caption{Photograph of the actual section of the anode frame with adjusting assembly. A wire carrier board is installed on the tensioning bar, and
     held in the correct position by the precision mounting pins. The tension is tuned by adjusting the bronze tensioning screws (only one shown in this
     image, but each threaded hole has a tensioning screw in the final
     installation). Note that there are also threaded holes for tensioning
     screws on the bottom of the C-channel frame, not visible in this image.
     }  \label{fig:tensionscrew}
  \end{center}
\end{figure}

The wire carrier boards attach to tensioning bars that are held inside a
C-channel frame, shown in the renderings of figure~\ref{fig:anodeframeCAD}. 
The exact positions of the tensioning bars are controlled by adjusting assemblies that allow a
small amount of pitch, roll, and yaw in each of the bars and many tensioning
screws (not shown in these figures). There are a total of 12 tensioning bar
segments, and these are grouped in four spans: two horizontal spans (top and
bottom of the anode frame) which are each constructed from 5 adjoining
segments, and two vertical spans (upstream and downstream sides of the anode
frame) which are each a single tensioning bar segment. There are 24 adjusting
assemblies; two per tensioning bar segment. The adjusting assemblies allow
fine-tuning during the installation and initial alignment of the wire planes,
but the bars are eventually fixed in place and tension on the wires is
applied by bronze tensioning screws, shown in figure~\ref{fig:tensionscrew}.

%\begin{figure}[h!]{}
%   \begin{center}{}
%   \begin{tabular}{c}{}
%     \includegraphics[scale=0.30]{TensionBar.png}
%     \end{tabular}
%     \caption{Photograph of a vertical MicroBooNE tensioning bar with precision 
%     mounting pins before wire carrier boards are installed.}
%     \label{fig:generalAlignment}
%  \end{center}
%\end{figure}

Before installing any wires, the tensioning bars must be squared and
aligned. This is done with the help of a team from the Fermilab metrology
group to adjust the bars within the C-channel frame such that the top and
bottom horizontal bars are parallel to each other, the upstream and
downstream vertical bars are parallel to each other, horizontal bars meet
vertical bars at right angles, and all four sides lie in a plane that is
parallel to the cathode plane. The angles of the four corners of the frame were all within 90.0$^{\circ} \pm$ 0.2$^{\circ}$ and the flatness of the frame surface was within $\pm$ 0.5~mm. The wire carrier boards are installed after
this alignment, and upon completion of the wire plane installation, tension
on the wires is tuned to the desired value by tightening the tensioning
screws to apply force on the tensioning bars. The tensioning process is
described in more detail in section~\ref{sec:tension}.

In preparation for installing the wires on the TPC, a trial installation
using sparsely populated wire carrier boards is undertaken. This trial
installation, known as the {\it mock installation}, is described in this
section, followed by a description of the full installation of real wires.

\subsection{Mock installation}\label{sec:mock}
The mock wire installation is designed to test the installation procedures
with a set of full-length prototype wires, allowing the installation team to
practice handling techniques and to identify areas in which procedures should
be modified before attempting to install the real set of 8256 wires.  This
trial run allows the team to detail the full installation and tensioning
procedure of the U, V, and Y wire planes. 

Wires for the mock installation are prepared using the wire winding machine
shown in figure~\ref{machine}. The only difference between the mock wires and
the real wires is the fact that the mock wires are entirely stainless steel,
while the real wires are stainless steel with copper coating and gold
flash. All other specifications are identical, and the mock wire carrier
boards are assembled according to the same procedures detailed in
section~\ref{winding}.

\begin{figure}[h!]{}
\begin{center}{}
\begin{tabular}{c}{}
\includegraphics[angle =0,width=10.5cm]{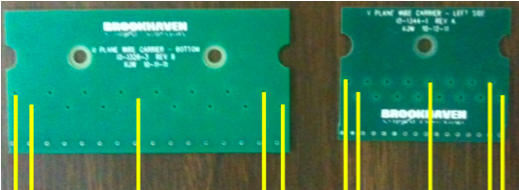}
\end{tabular}
\caption{Sketch of the number of wires and their locations on the wire boards 
in the mock wire installation.}
\label{mockfig}
\end{center}
\end{figure}

In order to reduce the total amount of time spent winding wires and producing
boards, only five wires are installed per wire carrier board, as shown
schematically in figure~\ref{mockfig}. These sparsely-populated wire carrier
boards allow the team to fully test the installation procedures for each of
the multiple wire planes (U, V, and Y) in the correct order and to understand
and test the wire tensioning procedure along the full length of the TPC. The
reduced number of wires per carrier board does not affect the installation or
tensioning procedures.

%\begin{figure}[tbp!]%RG

\begin{figure}[ht!]{}
\begin{center}{}
\begin{tabular}{cc}{}
\includegraphics[angle =0,width=7.5cm, height=10.8cm]{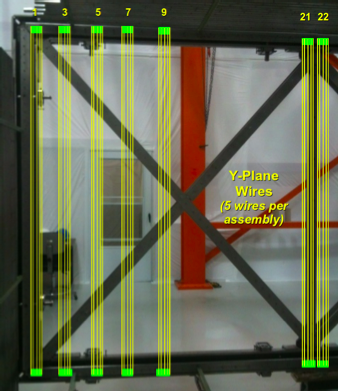}&
\includegraphics[angle =0,width=7.5cm]{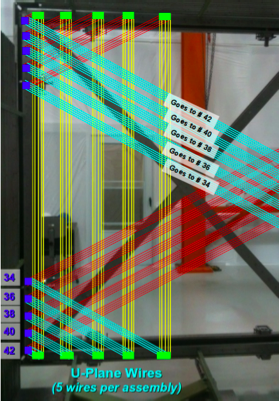}
\end{tabular}
\caption{Left: Y plane wire configuration for the mock-wire installation. Right: U plane wire configuration for the mock-wire installation (cyan) with previously installed Y plane (yellow) and V plane (red), for the ``left'' side of the TPC. Note that only 
the ``left'' side of the TPC anode frame is shown here. The ``right'' side of the 
TPC anode frame is populated with wire carrier boards according to the same prescription.}
\label{mockfig2}
\end{center}
\end{figure}

\begin{figure}[ht!]{}
\begin{center}{}
\begin{tabular}{c}{}
\includegraphics[angle =0,width=11.5cm]{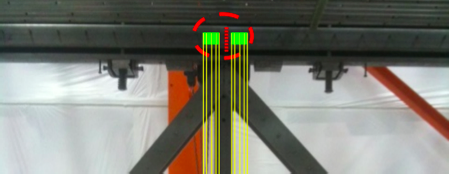}
\end{tabular}
\caption{A close-up of one section of the TPC anode frame where two horizontal 
tensioning bars join together. One Y-plane mock wire carrier board is installed 
on each side of the joint, corresponding to the two boards labeled as ``21'' and 
``22'' in figure}
%~\ref{mockfig2}. }
\label{mockfig3}
\end{center}
\end{figure}

Figure~\ref{mockfig2} (left) shows the layout of the Y plane boards for the mock
installation. The mock board installation locations are chosen to cover the
areas of the anode frame that are the most challenging to install (corners)
and the areas that are most critical for tensioning (joints), while covering
as large a fraction of the anode frame area as possible. Boards located near
the joints in the anode frame, where two stainless steel tensioning bars
adjoin, are an example of an area that is critical for consistent wire
tensioning. The vertical wire carrier boards labeled as ``21'' and ``22'' in
figure~\ref{mockfig2}, shown in more detail in figure~\ref{mockfig3}, are
installed to test how the joints behave and how the wire tensions on adjacent
boards change under tensioning in these regions. After the installation of the Y plane wires, the V and U planes wire carrier boards
stack onto the installed Y-plane boards as shown in figure ~\ref{mockfig2} (right).  The shortest
wires of the U and V plane are in the corners of
the anode frame. The longest wires, attaching to
the anode frame from one corner and extending to the
opposite corner (shown), are stacked on empty spacer boards to achieve the
correct plane-to-plane spacing. The main objective of installing these
particular wire carrier boards is to practice handling these 5-meter-long
wires.

The mock wire installation successfully tested the procedure and highlighted
steps that required special care; the quality assurance steps and
installation plan were modified to reflect the lessons learned here. We changed the initial position of the tensioning bars to facilitate the installation of the wires and we increased the number of people handling the boards. This
activity also trained installers with a technical rehearsal that ensured
smooth and coordinated operation during the full wire installation.

%Table \ref{mockfig5} summarizes the layout and locations of boards that were used in the mock-wire installation. A total of 58 boards, divided between U, V, Y planes, were installed and provide the installation team with invaluable information and expertise for the full wire installation. 

%\begin{figure}[ht!]{}
%\begin{center}{}
%\begin{tabular}{c}{}
%\includegraphics[angle =0,width=13.5cm]{mockTable.png}
%\end{tabular}
%\caption{List of all the wires boards that were used in the mock-wire installation. The spacer boards, used to complete the mock installation where all three planes are not installed, are not shown in this table.}
%\label{mockfig5}
%\end{center}
%\end{figure}

%\newpage
\subsection{Full installation}

%After the tensioning bars were in the proper initial position the entire structure was cleaned with alcohol to remove any possible dust, oil, or residues. Upon visual inspection after the alcohol rinse, some dust and small fibers (probably from the clean room suits) were found on the frame. The anode frame was then cleaned with dry air to remove any fibers or dust missed from the alcohol rinse. A second visual inspection was done and no fibers were found. A plastic cover was wrapped around the bottom field cage tubes to prevent dust or other debris to accumulate as shown in figure \ref{fig:flooring}. A temporary floor was installed over the field cage tubes to allow access in the TPC during the installation and to prevent any risk of damage to the tubes.\\
%
%\begin{figure}[ht!]{}
%\begin{center}{}
%\begin{tabular}{c}{}
%     \includegraphics[scale=0.11]{TPCPlastic.JPG}
%\end{tabular}
%          \caption{Picture of the lower field cage tubes covered in plastic to keep them clean during the wire installation process.
%     }  \label{fig:flooring}
%     \end{center}
%\end{figure}

The full set of wires are installed following the plan developed and tested
in the mock wire installation and took place inside of an enclosed assembly area outfitted with HEPA filter fans that provided a slight positive pressure relative to outside the assembly area. This provided a clean environment for the assembly, although the room did not have a clean room rating. After the initial alignment of the tensioning bars, the entire structure is cleaned with lint-free wipes and ethyl
alcohol. Then the wire carrier boards are installed serially, starting with
the Y plane, followed by the V plane, and finally the U plane. The wires on
each wire carrier board are visually inspected for damage before
installation. In the case that a damaged wire is found, the board is
partially disassembled, the damaged wire is replaced with a new wire, and the
board reassembled and tested again. Replacement of wires was only necessary
in a few cases where the wire sustained minor damage during installation. In
total, it took approximately one week to install all three planes of wires.

%?????
%Since the U and Y plane wires are designed to have applied bias voltages
%during detector operation, as described in section~\ref{planes}, leakage
%current tests of the capacitors on these wire carrier boards are performed
%just prior to installation. Leakage currents, if too large, will affect the
%readout signals from these wires. For the test, a bias voltage is applied to
%the wire carrier board and current is measured for each individual
%capacitor. Only boards with leakage currents below $1~\mu$A at their designed
%operating voltage (-220~V for U plane and +440~V for Y plane) are
%installed. V plane wire carrier boards are not tested, as they have no
%electrical components.

\begin{figure}[h!t]
   \begin{center}
   \begin{tabular}{c}
      \includegraphics[width=12cm]{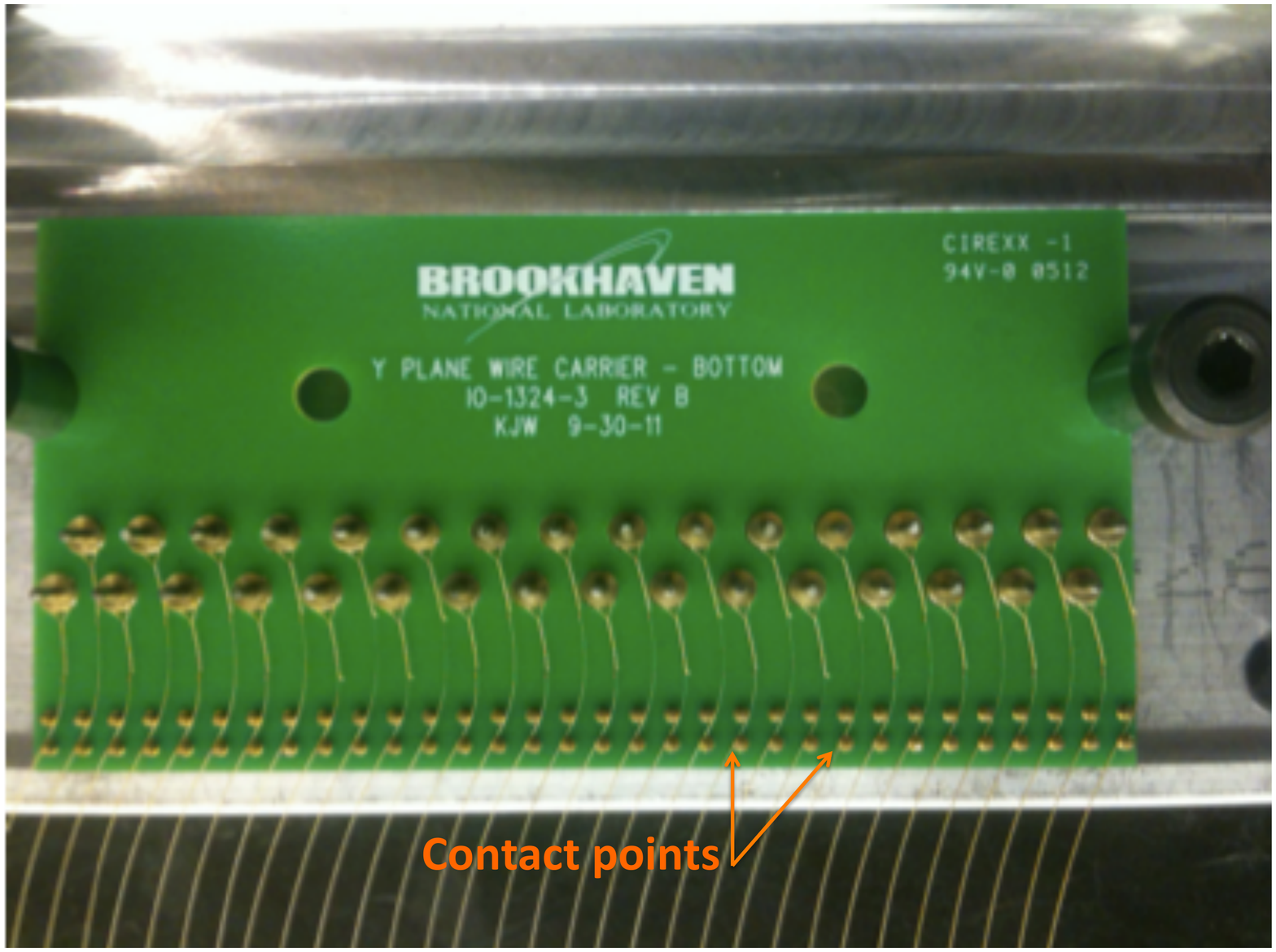}
     \end{tabular}
     \caption{Photograph of a wire carrier board showing the contact points
     between the board and the wires. This connection allows signal
     to be transmitted to the board electronics.  }  \label{fig:wirecontact}
  \end{center}
\end{figure}

\begin{figure}[h!t]
   \begin{center}
   \begin{tabular}{c}
      \includegraphics[width=15cm]{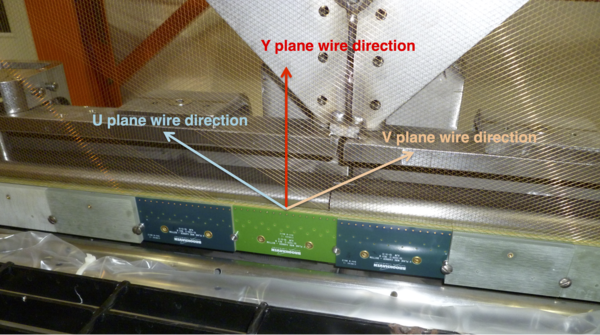}
     \end{tabular}
     \caption{Photograph with all three wire planes installed (wire
     orientations indicated by arrows) and showing the layout of the
     installed wire carrier boards on the frame. Three cover boards and one U
     plane wire carrier board are removed to better show the different layers.
     }  \label{fig:allwires}
  \end{center}
\end{figure}

Upon completing installation of each wire plane, a simple continuity test is
performed for each wire in that plane to ensure good electrical contact
between the wire and the carrier board pins, shown in
figure~\ref{fig:wirecontact}. Before beginning installation of the next
plane, the wires in the installed plane are also examined by eye, looking for
any kinked wires, abrasions, or other forms of damage. The final step, after
all wire planes are installed and inspected, is the addition of cover plates
that secure the stacked wire carrier boards to the tensioning bar, shown in
figure~\ref{fig:allwires}.

%To obtain proper alignment of all wire planes the jacking screws required several iterations of tuning. Cold welding between the stainless steel frame and jacking screws was a concern and was mitigated by using bronze screws, less likely to cold weld with stainless steel. In addition, both materials have a very similar linear thermal expansion coefficients. Before installing the MicroBooNE wires, the tensioning bars on the anode frame were aligned to an initial installation position using a theodolite and a laser level. This initial position ensured that during the installation process a wire would never be subject to an over tensioned state nor would it be loose enough to sag into another.\\%: $\alpha_{bronze} = 18.0 mm/K$ and $\alpha_{s.s.} = 17.3mm/K$.

\subsection{Wire tensioning}\label{sec:tension}

Proper tension of the wires is important for safe commissioning and
successful operation of the experiment. It is especially important to account
for wire tension during the initial cool-down phase of the experiment. During
this phase, the detector moves from a state where the cryostat is filled with
room temperature air to the state where it is filled with cold gaseous argon,
and although the wires and the TPC frame have similar thermal expansion
coefficients, the thin wires are subject to faster cool-down, and therefore
faster thermal contraction, than the rest of the massive stainless steel TPC
frame. If the wire contraction occurs much more quickly than the frame
contraction, wires could break or become permanently stretched. For this
reason, a thermal gradient limit is set during cool-down, requiring that the
temperatures at the top and bottom of the TPC frame never differ by more than
20~K. The nominal 0.7~kg (6.86~N) of tension established for the wires is
small enough to prevent wire breakage during cool down, but large enough to
limit the maximum wire sag due to gravity to less than 0.5~mm for any
5-meter-long U or V wire.

Tension is put on the wires by adjusting the bronze tensioning screws to push
on the tensioning bar, as shown in figure~\ref{fig:tensionscrew}. Since the
wires are grouped in sets of 16 or 32 on the wire carrier boards, and the
wire carrier boards are grouped together on adjoined rigid tensioning bars,
it is not possible to fine-tune the tension of individual wires. Instead, it
is necessary to adjust the positions of the tensioning bars to achieve as
uniform a map of wire tensions as possible across the full set of wire
carrier boards. 

\begin{figure}[ht!]{}
\begin{center}{}
\begin{tabular}{cc}{}
\includegraphics[scale=0.5]{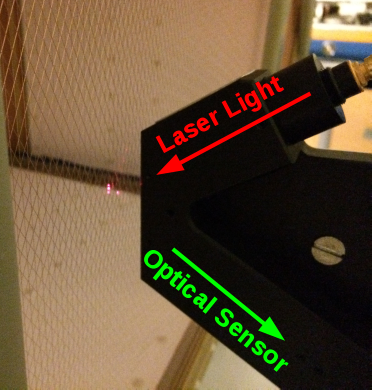}   
\includegraphics[scale=0.38]{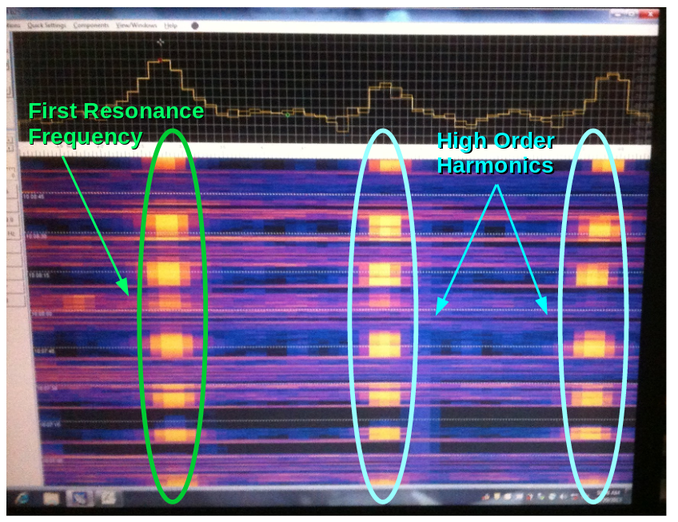}
\end{tabular}
\caption{Left: Photograph of the custom-designed apparatus used for measuring 
           wire tensions, which makes use of a focused LED and a
           photodiode. Right: Example output of the SpectrumAnalyzer software
           showing the resonant frequency (bright line on the left)
           and the higher order harmonics (lines in the middle and right). }
\label{fig:LaserTension}
\end{center}
\end{figure}

The tensioning process is necessarily iterative, as the adjustments in one
area of the TPC anode frame affect tensions in other areas due to the angled
U and V wire planes. It was realized during this process that the MicroBooNE design did not allow to adjust the tension of all the wires to the level of the original requirements, due to limitations of the tensioning system. This method also demonstrated that in the MicroBooNE design, it is not physically possible to measure the tension of all the wires because of mechanical constraints where the apparatus cannot access the wires. The tensions of approximately 95\% of the wires were
measured with a custom-designed apparatus designed and constructed by the 
Physical Sciences Lab of University of Wisconsin-Madison~\cite{PSL, device,device2}, shown in
figure~\ref{fig:LaserTension}. This device consists of an LED that is focused
onto a wire, and a photodiode that measures light reflected from the
wire. When the wire is lightly strummed, the output of the photodiode is fed
into a laptop with a spectrum analyzer that performs a fast Fourier transform
to convert the pulses into the frequency domain, and the peak frequency is
recorded for each board/wire measured. The frequencies measured can be
converted to tensions for known wire lengths.

%If a wire was over tensioned, it may break during the argon filling process. During the tensioning process, under-tensioning the wires were favored rather than over-tensioning them. The tensioning process was done by systematically adjusting the bronze tensioning screws across the entire frame until an acceptable overall frame tension is achieved.

%The tension of almost every wire was measured with a custom-designed testing apparatus that was produced for MicroBooNE, see figure \ref{fig:LaserTension}. A laser light was focused on a particular wire and the wire was carefully made to vibrate using a small plastic stick. A photodiode mounted inside of an optical lens measured the reflected light, which was converted to a frequency. The frequencies measured were read with a SpectrumAnalyzer \cite{SA} identifying the resonance frequency. The SpectrumAnalyzer also allowed the high order frequency harmonics to be seen, effectively allowing a more precise tension measurement as see in figure \ref{fig:LaserTension}.

Tension is measured using the characteristic resonance frequency of each
wire, which is related to its length ($L$ in $cm$), tension ($T$ in $N$), and
linear mass density ($\rho$, measured to be 0.00014~g/cm) through
equation~\ref{eq:1}.

\label{mathrefs}
\begin{equation}{} \label{eq:1}
f= \frac{1}{2 L}\sqrt{\frac{T}{\rho}}
\end{equation}

\begin{figure}[ht]{}
   \begin{center}{}
   \begin{tabular}{c}
   %\hspace*{-1.0cm} 
      \includegraphics[width=6.5in]{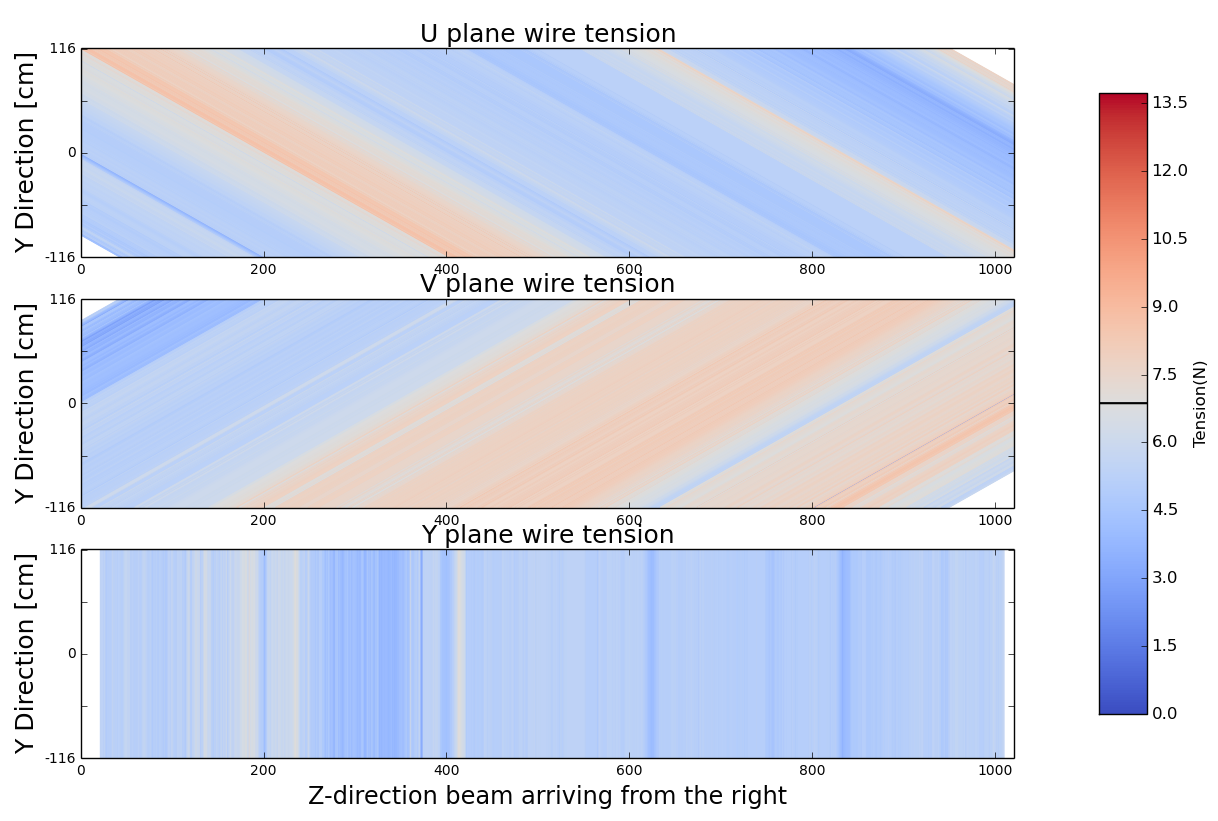}\\
      \hspace*{-1.0cm}
      \includegraphics[width=6.0in]{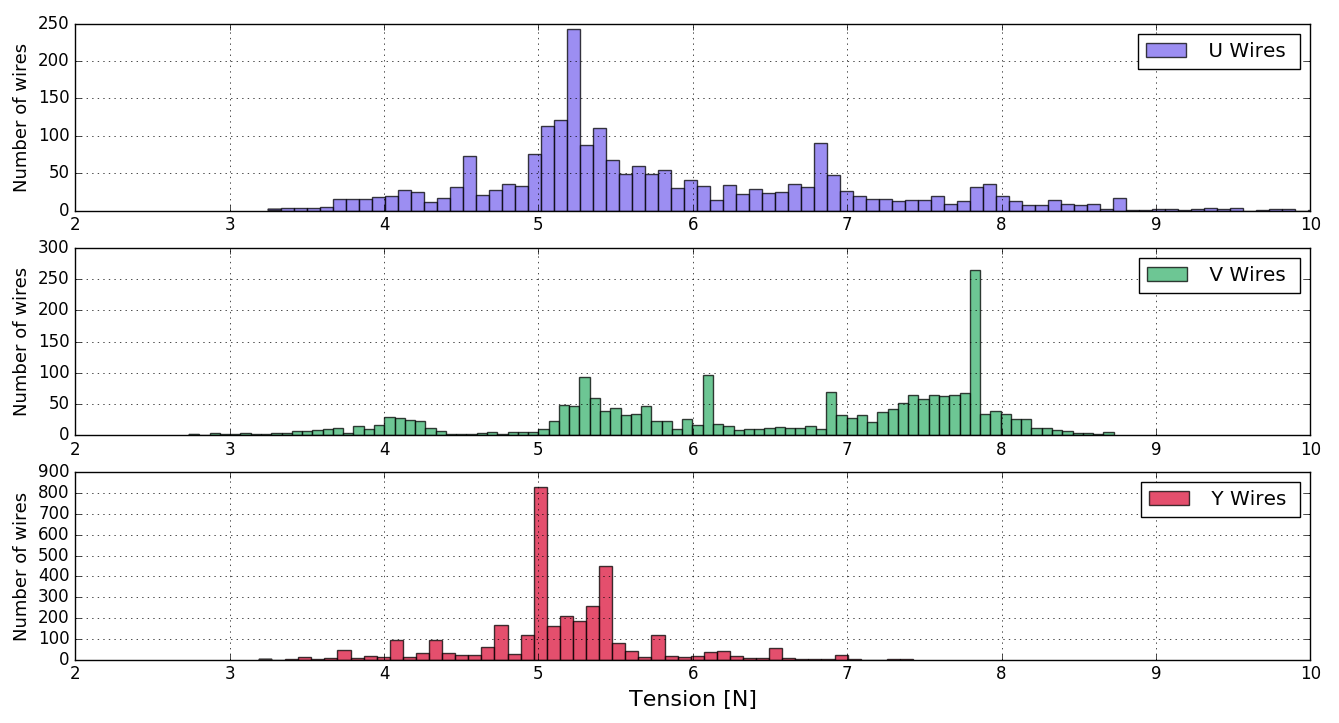}
    \end{tabular}
     \caption{Top:Tension map of all measured wires in each plane, as viewed from
     inside the TPC (beam enters from the right). The nominal tension (0.7~kg
     or 6.86~N) is marked on the legend with the black line and is
     represented by gray in the maps. Blue corresponds to wires with lower
     than nominal tension, and red to higher than nominal  tension. Bottom: Histograms of the measured tension values for the three wire planes.
    }  \label{fig:heatmap}
  \end{center}
\end{figure}

Figure~\ref{fig:heatmap} shows the tension measurement results for each of
the three planes. Tensions were measured for 2256 of the 2400 U wires, 2256
of the 2400 V wires, and 3312 of the 3456 Y wires, corresponding to 94.8\% of
the wires installed in the detector.

The overall average tensions of the U, V, and Y planes, respectively, are
5.78~N, 6.51~N, and 5.15~N. Table \ref{tension} also shows the values for the minimum and maximum measured tension values in each wire planes. Since it was physically impossible to adjust all the tensions as originally planned, lower-tension values were preferred to over-tension ones to
prevent wire breakage during cooldown, resulting in an overall average
slightly lower than the 6.86~N nominal value. These results can be related to the discussion in section \ref{tests}. The highest wire tension measured in each plane after the tensioning procedures is now 5.7/4.0/4.9 times lower than the lowest breakage tension value observed in the tests for the Y/U/V plane respectively at LN$_{2}$ temperature. This was considered satisfactory for MicroBooNE. We also looked at the expected sag for the wires with different tensions. The sag depends on the length of the wire as well as the tension applied to it. It was found that the largest possible sag was around 0.86~mm. This is slightly larger than the required 0.5~mm, but because of the mechanical constraints of the tensioning devices, this was judged satisfactory. This slightly larger sag value could affect the spatial resolution, however, it is mitigated by the redundancy of the three planes, allowing for at least two planes, where the sag is meeting the original requirements.  \\ 

\begin{table}[th!]
\label{tension}
\begin{center}
\begin{tabular}{|c|c|c|c|}
\hline
Plane  & Minimum tension (N) & Maximum tension (N)  & Average tension (N)\\
\hline
Y      & 3.19      &   7.43       &  5.15 \\
U      & 3.25      &   10.71      &  5.78  \\
V      & 2.74      &   8.73       &  6.51  \\
\hline
\end{tabular}
\caption{Tension values for the minimum, maximum and average tension for each wire planes.}
\end{center}
\end{table}

%\begin{figure}[H]{}
%   \begin{center}{}
%   \begin{tabular}{c}{}
%   \hspace*{-1.5cm} 
%      \includegraphics[width=6.4in]{WireTensionHist.png}
%     \end{tabular}
%     \caption{Distribution of individual wire tension for each of the three planes. The detector nominal tension was chosen to be around 0.7kg}
%  \label{fig:tensionhist}
%  \end{center}
%\end{figure}

%\clearpage
\section{Conclusion}\label{sec:x2}
The MicroBooNE TPC anode readout wire planes were assembled, installed, and
tested according to the procedures described in this document. The use of
semi-automatic wire winding machines considerably helped the preparation of
the individual wires. A strict quality assurance program was followed to
ensure that the wires were intact at every step of the process. This was
vital to the success of the plane construction, since a broken wire could
have had disastrous consequences on the experiment. Inspection of the wires 
inside the cryostat few months after the transport of the MicroBooNE cryostat 
also demonstrated the integrity of the planes ~\cite{wirecamera}. Several tests were
performed at different stages of the process to inform design decisions. It was found that tensioning all the individual wires to a desired value was unachievable with the tensioning bars and tension adjusting assemblies used for MicroBooNE.  The method described in this paper has some limitations but is appropriate for any medium-sized
LArTPC. For larger scale detectors, a more automated process should be
considered. The procedures described here can inform the design of future LArTPCs such as SBND \cite{sbnd} and DUNE \cite{dune}.

%\newpage % Please avoid layout-changing commands if not strictly necessary

%\begin{figure}[tbp] % figures (and tables) should go top or bottom of
                    % the page where they are first cited or in
\clearpage                    % subsequent pages

\acknowledgments

We would like to acknowledge the immense technical support from Tom Hurteau
as well as the help from undergraduate and graduate students for the wire
winding from Yale University: Kinga Partyka, Christina Brasco, Ben Elder and 
from Syracuse University: Thomas Badman, Deborah Noble, David Norcross. We gratefully acknowledge the Fermilab TRAC program, which provided funding for two of our authors. This paper is based 
upon work supported by the U.S. Department of Energy and the U.S. National Science Foundation. 
R. Guenette is now supported by the UK Science and Technology Facilities Council and A.M. Szelc 
is now supported by the UK Royal Society. Fermilab is operated by Fermi Research Alliance,
LLC under Contract No. DE-AC02-07CH11359 with the United States
Department of Energy.

\end{document}